# Positive algorithmic bias cannot stop fragmentation in homophilic networks

Chris Blex[ab] and Taha Yasseri[abc*]

[a]*Oxford Internet Institute, University of Oxford, Oxford, United Kingdom;*

[b]*The Alan Turing Institute, London, United Kingdom*

[c]*School of Sociology, University College Dublin, Ireland*

[*]Corresponding Author: Taha Yasseri: taha.yasseri@ucd.ie

Fragmentation, echo chambers, and their amelioration in social networks have been a growing concern in the academic and non-academic world. This paper shows how, under the assumption of homophily, echo chambers and fragmentation are system-immanent phenomena of highly flexible social networks, even under ideal conditions for heterogeneity. We achieve this by finding an analytical, network-based solution to the Schelling model and by proving that weak ties do not hinder the process. Furthermore, we derive that no level of positive algorithmic bias in form of rewiring is capable of preventing fragmentation and its effect on reducing the fragmentation speed is negligible.

Keywords: social networks, echo chambers, algorithmic bias, Schelling model, homophily

## 1. Introduction

Echo chambers or filter bubbles have been a concern since the early years of the commercial Internet, despite lacking a clear-cut definition. Sunstein (2017) warned that too much personalization may lead to 'online segregation', where individuals would surround themselves with people of ostensibly similar characteristics or ideologies. Van Alstyne and Brynjolfsson (2005) claim that the Internet makes it easier to find like-minded people and thus facilitates the creation of fringe communities that have a common ideology. These tendencies lead to the so-called Fragmentation Thesis (s. Bright, 2018) capturing the emergence of increasingly politically-driven divisions in online discussion networks. Fragmentation of (online) social networks has often been attributed to the sociological concept of homophily (Lazarsfeld and Merton, 1954; McPherson et al., 2001). Homophily states that individuals, which are similar on some sociological, economic, genetic or other plane are more likely to interact or form social ties. Homophily has a natural ally in Gratification Theory, a concept from psychology (Rosengren, 1974). It describes the selective behavior of individuals seeking to gratify certain pre-formed preferences in a high-choice environment, such as social media. Thus, social media platforms require individuals to be highly selective in their consumption of political content or engagement in political discussion. Given that these preferences are naturally homophilic, individuals are likely to be drawn to content and discussion partners aligned with their own beliefs, which in turn is encouraged by the affordances of a social media platform (Cho, 2003; Vaccari et al., 2016). Thus, social media might exacerbate an underlying sociological tendency of homogenous clusters within social networks.

Whilst clustering in online social networks is a well-established empirical fact (e.g. Adamic and Glance, 2005; Newman, 2006; Conover et al., 2011; Barbera, 2015a), a similarly interesting question is whether they are somewhat inevitable (Sasahara et al., 2019). The segregation of offline networks takes longer given geographical and temporal constraints and is thus prone to stochastic or social shocks upsetting the fragmentation process. However online social networks are much freer in their dynamics. From these considerations this paper derives its first research question:

**RQ1:** *Is fragmentation in (online) social networks inevitable, given presence of homophily and the high-pace dynamic of the structure of the network?*

Echo chambers are clearly a worrying phenomenon, since they are bad for social cohesion and democratic, deliberative decision-making processes (Sunstein, 2017; Bright, 2018). Algorithmically biased recommender systems have often been blamed for fostering these segregative tendencies. Thus, if part of the problem is algorithmic so should be its remedy. Therefore, this paper derives its second research question:

**RQ2:** *Can algorithmic bias counteract homophilic fragmentation of social networks? If so, what is the minimal algorithmic bias needed for fragmentation not to occur?*

This paper tries to answer these research questions by building and analytically solving a network representation of the Schelling model, which originally models spatial and residential segregation based on homophilic preferences of individuals (Schelling, 1971). Our model uses the benchmark case of an 'ideal liberal' network, i.e. a network in which individuals have no inherent biases in forming their connections (e.g. preferential attachment, social, political or racial biases inherent to the network structure, algorithmic biases) or limitations with whom and with how many they will forge a connection. This paper refers to this as the 'ideal liberal' case, since it corresponds to the ideal state of a liberal society and indeed the self-image of many social media platforms. Subsequently, the paper shows how even in networks with ideal conditions for heterogeneity a limited amount of homophily is sufficient to cause complete fragmentation in the limit. The paper supports this claim by proving that the result also obtains when including weak ties. Weak ties are secondary connections between individuals, i.e. "a friend of a friend" and have often been cited as having powerful hidden influences on social networks with regards to access to resources, information sharing, or network heterogeneity (Granovetter, 1973).

Lastly, the paper tries to derive the minimal algorithmic bias needed to stop the fragmentation of the network. This is done by rewiring edges between similar nodes to dissimilar nodes to increase heterogeneity. However, no level of rewiring bias can prevent the network from fragmenting and the influence on reducing the fragmentation speed is negligible. This poses some serious policy and ethical concerns for the accountability of administrators of social media sites. Firstly, it posits the question that if social media administrators are able to bias the tie creation process in favor of increased network heterogeneity, then should they be allowed to, obliged, or even litigated to do so. Secondly, if social media sites may be incapable of fostering sufficient heterogeneity to prevent unsustainable levels of polarization, the question begs whether they should be held accountable for political instability or social unrest. These are pressing ethical questions, since values such as individual freedom and privacy rights may conflict with wider political and social concerns of political uncertainty or discord.

This paper is organized in the following sections: Section 2 presents a review of related work, especially focusing on the Schelling model of segregation and the argument of weak ties posing a potential amelioration to the problem of network segregation. Section 3 presents the Homophily Theorem, which is a network operationalization of the Schelling model. Section 4 presents the Weakness of Weak Ties Theorem, which acts as an extension to this model. It shows how even when considering weak, i.e. secondary ties, the fragmentation process is uninhibited. Section 5 derives the minimal amount of algorithmic bias required for a network not to fragment. Section 6 summarizes the social science interpretations and implications of these findings.

**2. Literature Review**

Echo chambers or filter bubbles seem to lack a coherent definition or operationalization. One strand of studies defines echo chambers as highly-selective news diets (Gentzkow and Shapiro, 2011; Boxell et al., 2017; Dubois and Blank, 2018). Other studies have analyzed echo chambers and selective exposure based on media migration (Hollander, 2015) or measuring media diets based on URL-click data against a random baseline (Nikolov et al., 2015). Most studies on echo chambers rest on community detection and network clustering algorithms. Many of them have shown how networks seem to cluster along party lines or ideological beliefs on social media platforms (Adamic and Glance, 2005; Newman, 2006; Conover et al., 2011).

Many studies on echo chambers have tried to argue that echo chambers are overstated by pointing to the existence of weak ties (Bakshy et al., 2012; Barbera, 2015a, 2015b; Barbera et al., 2015; Hollander, 2015). Weak ties are supposed to counteract clustering movements by exposing users to opinions and information outside of their peer group. However, the probability of this being a decisive counteracting factor seems to be overstated if merely relying on the existence of weak ties. Iyengar and Hahn (2009) show that exposure to different views can actually entrench existing opinions even more, and Yardi and boyd (2010) claim that whilst exposure to opposing views might be high, meaningful engagement with them is low. Therefore, exposure to differing opinions may not be a sufficient metric to determine whether a society is polarized. Moreover, Marin (2012) shows that information holders in social networks base their decisions to share or withhold information on desire to help, reputational concerns, reluctance to appear intrusive, or fear of awkwardness resulting from negative consequences. This shows that when it comes to effective information sharing and information reception, weak ties alone are insufficient.

Another important notion seems to be bandwidth, which captures the level of trust, reputation, or usefulness that a neighboring node is associated with. Crudely speaking, a distant relative may be outside one's own echo chamber, but the effect of them posting controversial news stories on one's timeline is unlikely to affect one's opinion or the strength of the echo chamber effect. Crazy uncles rarely cause political epiphanies.

The underlying social force of echo chambers is homophily (Lazarsfeld and Merton, 1954; McPherson et al., 2001). Homophily is a well-documented sociological phenomenon (s. McPherson et al. for a comprehensive review) and has been frequently observed on social media (Barbera, 2015a, 2015b; Barbera et al., 2015; Nikolov et al., 2015; Tucker et al., 2018). Homophily is also the underlying principle of a famous model of spatial segregation, e.g. in housing. The renowned Schelling model (Schelling, 1971) shows how even limited homophilous preferences lead to high levels of segregation: Let there be a real line with dots of two different colors randomly assorted. Let each dot have a preference for its neighbors to be of a similar color. Should one dot move from a heterogenous neighborhood to a more homogenous neighborhood this trivially increases homogeneity in both neighborhoods. But also, other dots, which may have less homophilic tendencies are now proportionately surrounded by more dots of a different color, thus reinforcing their desire to move as well. It becomes clear, how the process unravels, and segregation becomes self-sustained, even if starting with very mild homophilic preferences and perfectly heterogenous neighborhoods.

Whilst, Schelling designed the original model on a line with nodes moving from one position to another in a one-dimensional space and the preferences driving segregation are hotly debated (Clark and Fossett, 2008), there has been a plethora of studies trying to generalize this model to more complex domains since. Vinkovic and Kirman (2006) adapt the model to cluster formation in physical systems, whilst Stauffer and Solomon (2007) compare residential separation to phase separation in physics. Some studies have extended the model to two- or multidimensional spaces in simulations and analytic proofs (Dall'Asta et al., 2008; Immorlica et al., 2015; Barmpalias et al., 2016, 2018). Studies more rooted in the model's home discipline of economics have included factors such as housing markets (Zhang, 2004) or used it in a game-theoretic study of collective action (Iwanaga and Namatame, 2013). Pancs et al. (2007) show how even if all actors have a strict preference for integration, the segregation result still holds.

The majority of studies introduce a tolerance parameter as a perturbation to the process and find that the segregation process can be stopped for specific values of such parameters (Gauvin et al., 2009; Gracia-Lazaro et al., 2011; Hazan and Randon-Furling, 2013; Immorlica et al., 2015; Sasahara et al., 2019). Immorlica et al. (2015) find an analytic result for an unperturbed one-dimensional Schelling model, starting from a random configuration, but finding a high probability of segregation. Some authors have translated the model into network environments (Banos, 2012). One study finds a network-based analytical solution to the Schelling model (Henry et al., 2011). Nodes rewire their edges stochastically based on different levels of aversion. The termination of an edge is based on attribute distance between two actors. The model finds a conversion to a distribution and derives a measure of attribute distance, measuring the degree of segregation. The intuitive result of the model is that segregation will always emerge outside of additional endogenous or exogenous drivers of network structure.

Translating the Schelling model from a real line to a network conjures some similarities and overlaps with the Axelrod model (Axelrod, 1997), which is a model of social influence and similarly to Schelling was based on moving around black and white dots on a checkerboard. Under the assumption that the more neighbors interact the more similar they become he shows how local convergence can lead to global polarization. With probability equal to their cultural similarity a randomly chosen site on the lattice will adopt one of the cultural features of a randomly chosen neighbors. Axelrod shows how a) the number of stable regions increases with number of possible traits and decreases with range of interaction, and b) the number of stable regions decreases with more cultural features and with large territories.

The model has sparked a number of extensions and combinations with the Schelling model. Guerra et al. (2010) translate the model from a lattice to a scale-free network with dynamic links and show that feature consensus is reached faster than global consensus. Lanchier and Scarlatos (2013) include homophily into Axelrod's model of social influence. That is, an interaction between two nodes persists until consensus is reached or one of the nodes terminates the tie and moves somewhere else. Thus, the model combines Schelling's idea of segregative homophily with Axelrod's social influence. The paper proves in a two-state model with arbitrary number of features that when the number of features exceeds the number of states per features, the Axelrod model leads to clusters.

Gracia-Lazaro et al. (2011) present another version of the Axelrod-Schelling model showing that a process of natural selection of advantageous traits seem to emerge. Rodriguez and Moreno (2010) include the mass media as an actor into the model as a node with more heavily weighted edges and a larger neighborhood and thus more influence. They show in a numerical simulation that the monocultural state is attained with stronger dependency on mass media strength and that network size seems to drive the system into a polarized society where all possible cultural configurations are present.

A recent paper by Sasahara et al. (2019) builds another hybrid Axelrod-Schelling model. They show how a network fragments and polarizes naturally from a non-polarized starting point and validate their model with simulations on Twitter data. Whilst using Schelling homophily for rewiring, the driving mechanism for them is bounded social influence. Similarly, Chitra and Musco (2019) build a hybrid model of social influence and homophily using the Friedkin-Johnson model (Friedkin and Johnsen, 1999). They include a network administrator into their model, who seeks to minimize disagreement between users thus fostering filter bubbles. In fact, changing total edge weights by 40 percent increases polarization 40-fold. They furthermore show that for a network generated by a Stochastic Block Model (SBM) that there may be low polarization and fragile consensus, which however is very easily perturbed leading to mass polarization. Yet, they also find that if the network administrator seeks to mitigate filter bubbles it only increases disagreement by five percent.

A related paper comes from Sirbu et al. (2019). The paper adapts the opinion dynamics model of bounded confidence by accounting for algorithmic bias. In original bounded confidence model (Deffuant et al., 2000), discussion partners are paired at random and converge in their opinion, if their original opinion difference is sufficiently small. In the converse case, they keep their opinion as it was before. Sirbu et al. (2019) modify this model by making the probability of two individuals interacting proportional to their opinion difference. That is, the smaller the opinion differential the more likely the interaction between two individuals. They show in a simulation that this leads to higher opinion fragmentation, increased polarization of opinions, as well as a stark decrease in the rate of opinion convergence.

**3. The Homophily Theorem**

*3.1 Model intuition*

A space is populated by individuals (nodes) of two types, e.g. red and green. The nodes are unconnected at the start of the model. In the 0th timestep they each form one connection to one other node with equal chance for a similar or dissimilar connection. The formation of these edges is unbiased, i.e. it is equally probable that a green node connects to another green node or a red node. From this timestep onwards, the edge formation is preferential. That is, at each timestep nodes form connections with other nodes, with a slight preference to connect to a node of similar color. The process is as follows: e.g. if a green node has a lot of connections to red nodes there is a high probability that the next connection it forms is to another green node. The more connections to dissimilar nodes, the higher the probability of forming further connections with similar nodes. This is an application of a Schelling model and follows a simple intuition: If a node prefers to be amongst similar nodes, but is currently surrounded by dissimilar nodes, it will have a stronger preference for the next connection to be to a node of a similar kind.

Note that for the model it is irrelevant whether these are strong or weak homophilic preferences. Additionally, nodes lose connections at each timestep with a certain probability. The longer a connection has persisted, the less likely a node is to lose this connection. I.e. the older the friendship, the more likely it is to persist. Letting this mechanism run its course, the network becomes entirely clustered into a green and red cluster in the limit. That is, the probability of forming a connection with a dissimilar node converges to zero.

*3.2 Mathematical formulation*

Let there be a network with an infinite number of nodes *n*, but no edges before time *t = 0*. At *t = 0* let each node form an edge with another node with no preferential bias with probability $p_t \sim Bernoulli(0.5, 0.5)$. Thus, there are $\frac{n(n-1)}{2}$ pairs of connected nodes.

From *t = 1*, at each time step each node creates a new connection at random.

Let $p_t \in [0,1]$ be the probability of one node making a connection to a node of the same type.

Let $q_t \in [0,1]$ be the probability of a node making a connection to node of a different type.

Trivially $p_t = 1 - q_t$.

Let $E_t \in \mathbb{Z}^+$ be the total number of edges with $E_t = E_{s,t} + E_{d,t}$, where $E_{s,t}$ is the number of edges between similar nodes and $E_{d,t}$ the number of edges between different nodes. Let $\mathbb{E}[E_{d,i,t}]$ be the expected number of edges from node $i$ to nodes of a different type to $i$.

$\mathbb{E}[E_{d,i,t}] = q_{i,t} + k_d E_{d,i,t-1}$, where $k_d$ is the proportion of nodes of a different type from the previous period, that are still attached to node $i$, i.e. the proportion of different connections kept. This process can be conveniently described by a hazard function, i.e. a decay function. The latter captures that the older a connection gets, the less likely it is to be severed, since older connections tend to be socially stronger.

Iterating this equation backwards yields

$$\mathbb{E}[E_{d,i,t}] = q_{i,t} + \sum_{j=0}^{t} k_{d,t-j} q_{i,t-1-j}$$

with $E_0 = \frac{n(n-1)}{2}$ and $E_{d,0} = \frac{n(n-1)}{4}$. N.B. that the process for the expected number of similar edges looks the same, albeit with $k_d < k_s$.

For all $t > 0$ let $p_t = f(\frac{E_{d,t-1}}{E_{t\_1}})$, where $f' > 0$ and $f'' < 0$ with $f(1) = 1$,. That is, the probability increases concavely in the number of edges formed to dissimilar nodes with a maximum reached at 1. The latter implies that individuals have no saturation point for homophily, something that can also be found in the Schelling model.

Using the mean-field approximation we can assume $p_t = f(\frac{\mathbb{E}[E_{d,t-1}]}{E_{t\_1}})$.

**Theorem 1:** For any concave strictly monotonically increasing function $f : [0,1] \rightarrow [0,1]$, $p_t$ converges to 1 and thus $\mathbb{E}[E_{d,t}]$ converges to zero.

*3.3 Proof*

Consider the following system of equations:

$$p_t = 1 - q_t = f\left(\frac{\mathbb{E}[E_{d,t-1}]}{E_{t-1}}\right) \quad (1)$$

$$\mathbb{E}[E_{d,i,t}] = q_{i,t} + \sum_{j=0}^{t} k_{d,t-j} q_{i,t-1-j} \quad (2)$$

where $f' > 0$ and $f'' < 0$

The derivative of the probability of making similar connections with respect to the probability of making dissimilar connections is given by:

$$\frac{\partial}{\partial q_{i,t-1}} p_{i,t} = f'\left(\frac{\mathbb{E}[E_{d,t-1}]}{E_{t-1}}\right) t^{-1} > 0 \quad (3)$$

This implies that as $t \to \infty$, $\frac{\partial}{\partial q_{i,t-1}} p_{i,t} \to 0$. And thus for all strictly monotonically increasing concave functions $f$ on $[0,1]$, $p_{i,t} \to 1$, implying $q_{i,t} \to 0$ and $\mathbb{E}[E_{d,i,t}] \to 0$.

The proof above shows how social networks of infinite size with limited homophily will gravitate towards complete fragmentation, even under ideal conditions for heterogeneity given a presence of limited homophily. Given no exact functional form other than an operationalisation of Schelling homophily was imposed on $p_{i,t}$, the model can incorporate other operationalisations of homophily, e.g. attribute vectors. It furthermore proves the Schelling result of segregation in a network setting.

**4. The weakness of weak ties theorem**

This theorem will act as an extension to the *Homophily Theorem*. It will add a form of resistance to the fragmentation based on weak ties of one group to another. Weak ties have been extensively discussed in the literature, since Granovetter's seminal paper on how weak ties are often decisive in resource and information extraction (Granovetter and Soongt, 2016). In social media analyses weak ties have often been cited as counteracting mechanisms to the existence of echo chambers (Barbera, 2015a, 2015b; Barbera et al., 2015; Hollander, 2015; Bakshy et al., 2015). The *Weakness of Weak Ties Theorem* will attempt to show that the result of the *Homophily Theorem* still holds even with the existence of weak ties.

*4.1 Mathematical formulation*

At every time step, in addition to the new random connection (primary) each node connects to a neighbor of its neighbors who is dissimilar to itself (secondary). Let node $i$ be attached to a similar node $k$ with probability $p_i$. Node $k$ will be connected to a dissimilar node with probability $q_k = 1 - p_k$. Thus, the expected number of dissimilar ties, primary and secondary (i.e. weak), for node $i$ are given by the following expression

$$\mathbb{E}[E_{d,i,t}] = q_{i,t} + q_{i,t}p_{k,t} + p_{i,t}q_{k,t} + k_d E_{d,i,t-1}$$

After substituting the expression for $E_{d,k,t}$, iterating the equation backwards and considering it in continuous time:

$$\mathbb{E}[E_{d,i,t}] = q_{i,t} + q_{i,t}p_{k,t} + p_{i,t}q_{k,t} + \sum_{j=0}^{t} k_{d,t-j}[q_{i,t-j-1} + q_{i,t}p_{k,t-j-1} + p_{i,t-j-1}q_{k,t-j-1}]$$

**Theorem 2** For any concave strictly monotonically increasing function $f : [0,1] \to [0,1]$, $p_t$ converges to 1 and thus $\mathbb{E}[E_{d,t}]$ converges to zero, even in the presence of secondary ties.

*4.2 Proof*

The derivative of equation (3) has now changed to

$$\frac{\partial}{\partial q_{i,t-1}} p_{i,t} = f'\left(\frac{\mathbb{E}[E_{d,t-1}]}{E_{t-1}}\right) \frac{1 + p_{k,t-1}}{t} = f'\left(\frac{\mathbb{E}[E_{d,t-1}]}{E_{t-1}}\right) \frac{1 + (1 - q_{k,t-1})}{t} > 0$$

Given that $p_{k,t}$ is bounded in [0,1] for all $t$, this means that the derivative converges to zero. Note that the convergence rate is being slowed down compared to the result of the *Homophily Theorem,* but that this deceleration quickly declines over time as $t$ increases and by symmetry $q_{k,t-1}$ decreases. Thus, for all strictly monotonically increasing functions $f$ on [0,1], $p_{i,t} \to 1$, implying $q_{i,t} \to 0$ and $\mathbb{E}[E_{d,i,t}] \to 0$, even in the presence of weak ties.

Thus, the *Homophily Theorem* even persists in the presence of secondary, or weak ties. Note that the convergence time inconsiderably, but technically increases for the likelihood of forming secondary ties. Yet, the increase in convergence time is practically negligible. This indicates a robustness of the *Homophily Theorem* and acts as a warning that weak ties may not be enough to hedge against echo chambers.

## 5. Algorithmic bias

Now we investigate the possibilities for stopping the fragmentation by finding the minimal algorithmic bias that produces a non-zero value for the expected number of different ties $\mathbb{E}[E_{d,i,t}]$. This will be done in form of a *rewiring bias*. The rewiring bias simply states that a certain proportion of existing similar connections will be changed to dissimilar connections through an algorithmic mechanism, here given by $\phi_{i,t}$, thus increasing the heterogeneity in the network. The intuition behind this can be described as individuals being shown more content contrary to their views than they normally would be exposed to given their network. I.e. a red node with a majority of red connections will see more 'green' content than they normally would. This is operationalized as follows:

$$\mathbb{E}[E_{d,i,t}] = q_{i,t} + \sum_{j=0}^{t} k_{d,t-j} q_{i,t-1-j} + \sum_{j=0}^{t} k_{b,t-j} \sum_{l=1}^{t} (\phi_{i,t-l} - k_{s,t-l})(t - l - 1 - q_{i,t-l-1}) + (\phi_{i,t} - k_{s,t})(t - 1 - q_{i,t-1}),$$

with $k_d < k_b < k_s$ and $k$ being assumed to be a Weibull process $k = 1 - \lambda^{-\gamma} \gamma t^{\gamma-1}$, where $\lambda \geq 1$ is a scale parameter and $\gamma < 1$.

Note, that the result remains unchanged for homogenous *k* across all types of connections and the distinction between biased, dissimilar, and similar connection is done here for potential extensions of the model and rigor. Another assumption for the rewiring bias is that the longer a connection has persisted, the more difficult it becomes to rewire it. If I have been a content consumer of BBC News the likelihood of accepting an imposed change to Fox News is rather low and vice versa.

Consider the equation above with all variables at their steady-state value, denoted here by an *. N.B. that *k* in all cases converges to one for large *t*.

$$\mathbb{E}[E_{d,i,t^*}] = q^* + t^*q^* + t^*(t^* - 1)(\phi^* - 1)(t^* - 2 - q^*) + (\phi^* - 1)(t^* - 1 - q^*)$$

Since $\phi$ is bounded on [0,1] for all *t*, the optimal and in this case minimal level of bias is the boundary solution $\phi^* = 1$. This implies that every similar connection is changed to a dissimilar connection. Consequently,

$$\mathbb{E}[E_{d,i,t^*}] = (1 + t^*)q^*$$

And

$$\frac{\mathbb{E}[E_{d,i,t^*}]}{\mathbb{E}[E_{s,i,t^*}]} = \frac{(1 + t^*)q^*}{(1 + t^*)p^*} = \frac{q^*}{p^*}$$

For the probability of making similar connections in the steady-state this implies:

$$p^* = f\left(\frac{\mathbb{E}[E_{d,i,t-1^*}]}{t^*}\right),$$

which in the limit converges to

$$p^* = f(1) \text{ for } q_{t^*-1} > 0$$

The latter implies $p^* = 1$ and therefore $q^* = 0$ and $\mathbb{E}[E_{d,i,t^*}] = 0$. Thus, for any $\phi^* < 1$, $\mathbb{E}[E_{d,i,t^*}]$ diverges and in fact becomes negative, whilst for $\phi^* = 0$ $\mathbb{E}[E_{d,i,t^*}]$ converges to zero. $\phi^* > 1$ implies that there is more rewiring of similar connections to dissimilar ones than actually exist, which is a contradiction. Therefore, no rewiring bias can stop fragmentation in the limit. The homophilic powers are simply too strong to prevent fragmentation even with severe intervention, in this case rewiring all dissimilar nodes of an individual to similar ones.

Furthermore, consider the influence of the algorithmic bias on the speed of convergence. Equation (3) changes under algorithmic bias to

$$\frac{\partial}{\partial q_{i,t-1}} p_{i,t} = f' \frac{1 - (\phi_{i,t} - k_{s,t})}{t - 1}$$

Note that as $t \to \infty$, $k_{s,t} \to 1$ and therefore the derivative converges to

$$\frac{\partial}{\partial q_{i,t-1}} p_{i,t} = f' \frac{\phi_{i,t}}{t - 1}$$

Similar to the *Weakness of Weak Ties Theorem* the process can be slowed down using the bias $\phi$. However, the necessary bias to prevent $p_{i,t}$ from approaching one increases in *t* and therefore necessarily diverges, even if a bias bigger than one would be conceptually possible.

## 6. Discussion

The *Homophily Theorem* is a form of a stylized Impossibility Theorem. That means that even in ideal, stylized conditions, in social networks with any degree of homophily, heterogeneity is impossible. The type of homophily or operationalization is irrelevant since the function $p_t$ is as general a formulation of homophily as it could possibly be. Whilst the assumption and operationalization seem abstract and unrealistic at first, it is worth noting that the model is thus capable of a number of different interpretations. Edges can be interpreted as friendship ties, or (more realistically) as interactions in a discussion or sharing network. Thus, the severance of ties can simply mean two users stopping to talk to each other rather than the more drastic social action of unfriending. Furthermore, a severance of ties can also represent that the tie simply has become irrelevant. Whilst it may exist *pro forma* in shape of a digital connection, the lack of interaction renders it meaningless. Lastly, the Weibull process can alternatively be thought of as the strength of a tie diminishing rather than the probability of it vanishing.

*The Weakness of Weak Ties Theorem* makes the results of the *Homophily Theorem* more robust. It shows that even weak ties do not only not hedge against fragmentation, but do not even slow down the process. It is however important to mention that the model omits certain social factors which may drive network unification, such as a common goal, the rule of law, or cultural similarities. One such example can be found in Torok et al. (2013) and Iñiguez et al. (2014), where the collaborative project of a Wikipedia article and thus indirect interactions between agents increases consensus in a network over time.

Notwithstanding, it can be shown that homophily massively increases in scale and scope in social networks with high flexibility of node attachment and detachment, such as online social networks. It is this flexibility generated by a site's affordances that facilitates users to gratify their desires, such as homophily quickly and almost costlessly. Therefore, fragmentation becomes a natural point of gravitation for these networks. Whilst complete fragmentation is obviously unrealistic, it is important to see that it is a natural tendency for these types of networks. It is worth noting that the main assumption of this model is obviously that fragmentation is driven by homophily and that it ignores social influence. The question which of the two prevails more in e.g., opinion networks on social media has not been empirically answered yet. Should opinions be sufficiently robust on social media and confirmation biases sufficiently strong, it seems more realistic that homophily is the driving force of tie generation and attrition in opinion networks. Furthermore, Iyengar and Hahn (2009) show that exposure to different views can actually entrench existing opinions even more, i.e., social influence may also worsen rather than ameliorating the problem.

The introduction of algorithmic bias into the model shows that a network administrator may have no substantial tools to alleviate this.. Opposite to the main idea of Chitra and Musco (2019) in this model the network administrator has no incentive to minimize disagreement and seeks to stop fragmentation. The model shows that in order to do so, the administrator has to especially counteract the severance of ties by rewiring. This could for instance mean that old connections, which have become irrelevant are now sought to be reinvigorated, e.g. by recommender systems pushing them back to the top of the feed. However, Sasahara et al. (2019) point out that recommending things that will be ignored is not a good strategy. They propose that preventing triadic closure would be an even more neutral interference. Yet, they also agree that complete severing of ties should be discouraged (e.g. by the inclusion of affordances such as snoozing).

The most concerning result is that even without making any assumptions about the functional form of algorithmic rewiring bias and rewiring every similar connection, the network administrator is incapable of preventing the network to fragment in the limit. Whilst in the early stages of tie generation the use of algorithmic bias might slow down the fragmentation process, even this becomes impossible very quickly over time. This furthermore suggests that complete fragmentation could only be fought by additive bias, which is very invasive and might also not be accepted by users. Moreover, this poses big ethical and regulatory questions on whether and how network administrators can or should interfere. It furthermore may indicate that the social forces of fragmentation in such highly flexible networks can only be tentatively mitigated by network-based or algorithmic mechanisms.

This paper suggests how echo chambers are seemingly inevitable in social networks with high flexibility of node attachment and detachment. Even in ideal conditions for heterogeneity (such as no network inherent biases for tie creation or algorithmic bias) a limited amount of homophily is sufficient to cause the network to fragment in the limit. Furthermore, this paper shows that even the introduction of the highest possible degree of algorithmic bias in form of rewiring is insufficient to prevent network fragmentation in the limit. Whilst it may slow down the process in its early stages, fragmentation quickly becomes an insurmountable problem. Thus, there are major regulatory and ethical questions concerning the role of network administrators. Firstly, if network-based mechanisms such as biased recommender system can barely mitigate the problem, what is the responsibility of network administrators and social media sites (especially now that Pandora's box has been opened)? Secondly, how much should network administrators intervene in this process and how much does such intervention warrant regulation?

On a technical level the paper produces a straightforward translation of the Schelling model into a network-based model. Furthermore, whilst the Schelling model has been shown as robust in simulations it hitherto has had only one analytical solution (however not in network form). This is a considerable methodological contribution, since as Brandt et al. (2012) point out that the Schelling model is "surprisingly difficult to analytically prove or even rigorously define the segregation phenomenon observed qualitatively in simulations". Another contribution is the result that weak ties do not ameliorate the fragmentation process. Even stronger, weak ties do not even substantially slow down fragmentation. This is a warning call, since there is a prevalent view in the literature that weak ties hedge against the building of echo chambers (Bakshy et al., 2012; Barbera, 2015a, 2015b; Barbera et al., 2015; Hollander, 2015).

Future work could be conducted in testing the fundamental assumption of the model, namely homophily in opinion networks. That is, are dynamics in opinion networks on social media mainly driven by homophily, due to psychological phenomena like confirmation bias, or are they sufficiently malleable by neighboring nodes. The model could be extended to incorporate such social influences. Another extension could be proving that the model holds in arbitrary dimensions of or different operationalization of homophily. Finally, the behaviour of the model in finite networks and under additive bias could be explored.

**Declaration of interest statement**
No financial interest or benefit has arisen from the direct applications of this research.

**Acknowledgements**

The authors would like to thank Oliver M. Crook for his advice on using Euler-Lagrange equations for finding the minimal algorithmic bias. The authors would also like to thank Renaud Lambiotte for his advice.

**Funding details**

Chris Blex was funded by the Alan Turing Institute Studentship. Taha Yasseri was partially supported by The Alan Turing Institute under the EPSRC grant EP/N510129/1. The Funder had no role in the conceptualization, design, data collection, analysis, decision to publish, or preparation of the manuscript.